\documentclass[
twocolumn,
]{ceurart}

\usepackage{listings}
\usepackage{booktabs}
\usepackage{multirow}
\usepackage{listings} 
\usepackage{graphicx}
\usepackage{array}
\usepackage{lipsum} 
\usepackage{svg}
\usepackage{float} 
\usepackage{placeins} 
\usepackage{algorithm}
\usepackage{algorithmic}
\usepackage{caption}
\usepackage{subfigure}
\usepackage{subcaption}
\begin{document}

\copyrightyear{2024}
\copyrightclause{Copyright for this paper by its authors.
  Use permitted under Creative Commons License Attribution 4.0
  International (CC BY 4.0).}

\conference{The 1st Workshop on Risks, Opportunities, and Evaluation of Generative Models in Recommender Systems (ROEGEN@RecSys 2024), October 2024, Bari, Italy.}

\title{Elaborative Subtopic Query Reformulation for Broad and Indirect Queries in Travel Destination Recommendation}


\author[1]{Qianfeng Wen}
\fnmark[1]
\address[1]{University of Toronto, Toronto, Canada}

\author[1]{Yifan Liu}
\fnmark[1]

\author[1]{Joshua Zhang}

\author[1]{George Saad}

\author[1]{Anton Korikov}

\author[2]{Yury Sambale}
\address[2]{Air Canada, Montreal, Canada}

\author[1]{Scott Sanner}[
email=ssanner@mie.utoronto.ca
]
\cormark[1]

\fntext[1]{These authors contributed equally.}
\cortext[1]{Corresponding author.}

\begin{abstract}
In Query-driven Travel Recommender Systems (RSs), it is crucial to understand the user intent behind challenging natural language (NL) destination queries such as the broadly worded ``\textit{youth-friendly activities}'' or the indirect description ``\textit{a high school graduation trip}''. Such queries are challenging due to the wide scope and subtlety of potential user intents that confound the ability of retrieval methods to infer relevant destinations from available textual descriptions such as WikiVoyage.
While query reformulation (QR) has proven effective in enhancing retrieval by addressing user intent, existing QR methods tend to focus \textit{only} on expanding the range of potentially matching query subtopics (breadth) or elaborating on the potential meaning of a query (depth), but not both.  
In this paper, we introduce \textbf{E}laborative Subtopic \textbf{Q}uery \textbf{R}eformulation (\textbf{EQR}), a large language model-based QR method that combines \textit{both} breadth and depth by generating potential query subtopics with information-rich elaborations.
We also release \textbf{TravelDest}, a novel dataset for query-driven travel destination RSs. Experiments on \textbf{TravelDest} show that \textbf{EQR} achieves significant improvements in recall and precision over existing state-of-the-art QR methods. 
\end{abstract}

\begin{keywords}
    Recommender Systems \sep
    Query Reformulation \sep
    Large Language Models \sep
    Travel
\end{keywords}

\maketitle

\section{Introduction}
\label{sec:intro}

\begin{figure*}[t!]
    \centering   
    \includegraphics[width=1.0\textwidth]{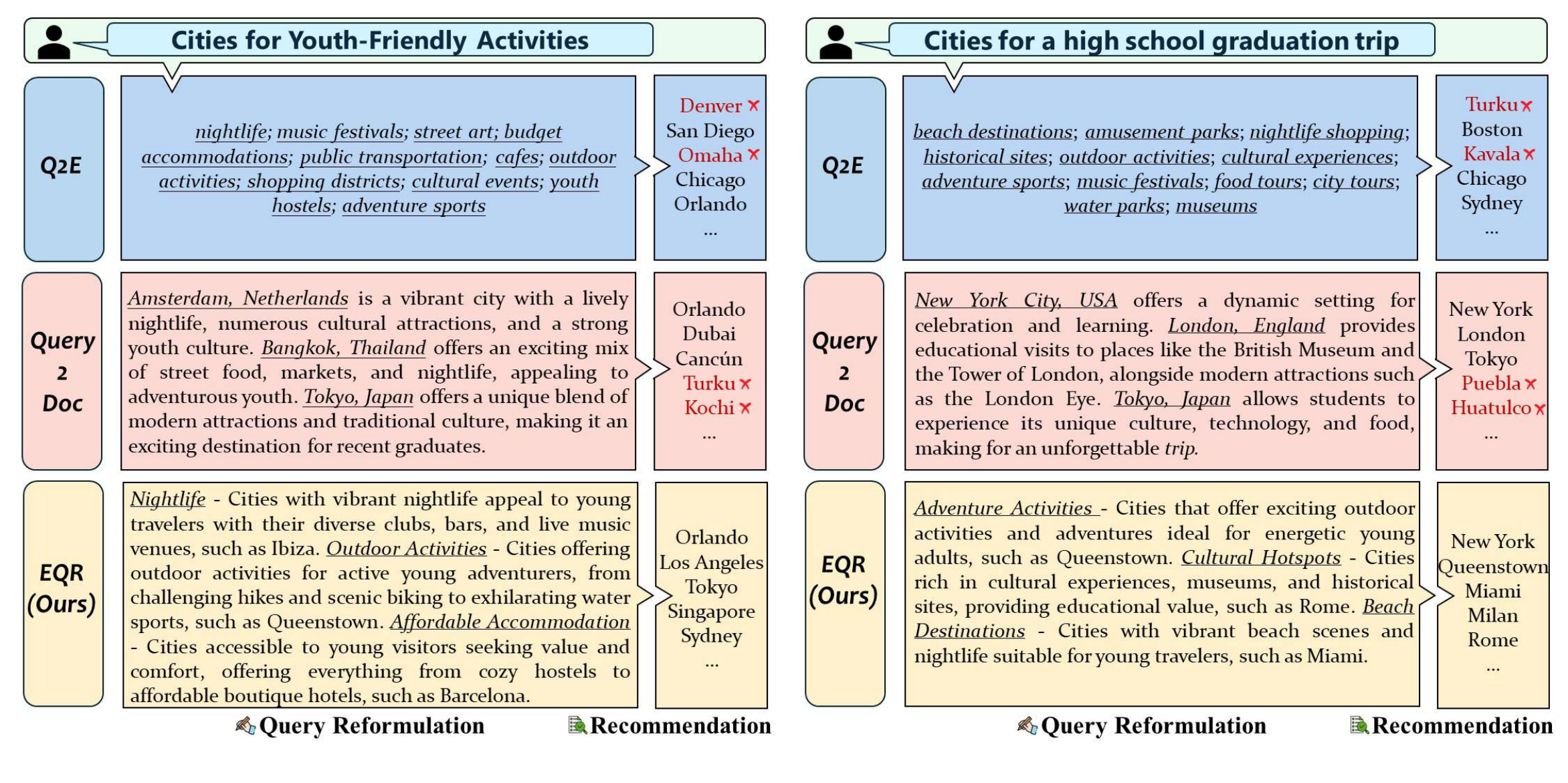}
    \caption{Examples of broad (\textit{Cities for youth-friendly activities}, right) and indirect (\textit{Cities for a high school graduation trip}, left) queries. The comparisons are made among different QR methods discussed in Section \ref{qr}: Q2E \cite{q2dq2e} (top), Query2Doc \cite{query2doc} (middle), and EQR (bottom). The top 5 recommendations are shown, with irrelevant results marked in red with an {\color{red} $X$}.  One can clearly identify that \textbf{Q2E} covers subtopic breadth, but lacks depth of description, while \textbf{Query2Doc} provides in-depth concrete suggestions without covering general subtopics.  \textbf{EQR} covers both general subtopic breadth with deep subtopic elaboration.}
    \label{fig:examples}
\end{figure*}

Air travel is a rapidly growing industry with revenues exceeding 800 billion USD in 2023 \cite{driversairline}. Providing personalized travel destination recommendations to users based on their natural language (NL) queries offers novel opportunities for the air travel industry to engage potential travelers and better serve their travel needs. 


A Query-driven Travel Destination Recommender System (RS) can leverage a retrieval subsystem to \emph{match} NL queries to relevant passages in destination descriptions (e.g., sourced from CC-licensed content such as WikiVoyage\footnote{\label{fn:wikivoyage}\url{https://www.wikivoyage.org/}}) and \emph{score} destinations (e.g., as later defined in Algorithm~\ref{alg:dest_score}).  However, such a retrieval approach is prone to fail in light of challenging NL queries that often occur in the Travel RS setting as evident in the following two query types:
\begin{enumerate}
    \item \textbf{Broad Queries}: These are queries that specify a broad category and imply multiple potentially relevant subtopics (e.g., \textit{``cities for youth-friendly activities''}). 
    \item \textbf{Indirect Queries}: These are queries that do not directly reflect the user intent but require several further reasoning steps (e.g., \textit{``cities for a high school graduation trip''}).

\end{enumerate}

To address such challenging NL queries, existing retrieval research has explored Query Reformulation (QR) to improve intent understanding~\cite{InferringQueryIntentfromReformulationsandClicks, AutomaticQueryExpansion} with recent methods additionally leveraging Large Language Models (LLMs)~\cite{languagelearningAreFewShotLearners, 2023GenQR}. 
LLM QR methods focus on either (a) 
expanding queries w.r.t.\ diverse keywords using LLMs, which can be interpreted as a focus on adding subtopic \textit{breadth} to the original query~\cite{q2dq2e, genqr} or (b) 
generating paraphrases or relevant answer passages, which can be interpreted as a focus on adding conceptual \textit{depth} to the description~\cite{hyde, q2dq2e, query2doc, gqrgqe, 2023GenQR}.


We conjecture that \textit{both} breadth and depth are key for connecting a user's NL queries to relevant recommendations in travel destination RSs.  Further, we observe that LLMs' general reasoning abilities \cite{tafjord2020proofwriter, yao2024tree} allow them to expand queries with \textit{both} diverse (i.e., breadth) and nuanced (i.e. depth) content that facilitates more effective sparse and dense retrieval with broad and indirect queries.  

In summary, we make the following contributions: 

\begin{enumerate}
    \item We propose a novel LLM-based \textbf{E}laborative Subtopic \textbf{Q}uery \textbf{R}eformulation (\textbf{EQR}) method \footnote{code: \url{https://github.com/YifanLiu2/ROEGEN-RecSys-24-EQR.git}}
\footnote{video:\hspace{2pt}\url{https://youtu.be/WoU77Nw_Z8o}}
 that infers multiple subtopic intents covering an original query (i.e., breadth) while providing information-rich elaborations of each subtopic (i.e., depth).  \textbf{EQR} 
    leverages the abilities of LLMs to understand and reason about broad and indirect user intent on the query side that facilitates effective matching via sparse and dense retrieval.  A comparison of two existing QR methodologies with \textbf{EQR} in Figure~\ref{fig:examples} provides clear evidence that \textbf{EQR} better addresses breadth and depth as we have conjectured. 
    \item We introduce and release \textbf{TravelDest}, a novel manually curated benchmark dataset for query-driven travel destination RSs with 50 \textit{broad} and \textit{indirect} NL queries and complete relevance labels for 774 destination cities to support our initial comparative evaluation of Travel RS QR methods.
\end{enumerate}
We conclude with a comparative evaluation of QR methods on \textbf{TravelDest} and find that \textbf{EQR} outperforms existing methods in terms of recall and precision metrics, thus improving the ability of Travel RSs to address challenging broad and indirect NL queries. 

\section{Related Work}



We review Travel Recommender Systems followed by Query Reformulation that is the methodological focus of our contributions. 

\subsection{Travel Recommender Systems}
Traditional travel RSs have primarily focused on query-free recommendations using collaborative filtering (CF) that leverages the collective travel patterns of users
\cite{aggarwal2016recommender, sarkar2023tourism}, or content-based filtering (CBF) that 
recommends items similar to 
a user's past preferences using 
item features such as location, cost, and reviews \cite{lops2011content, poi, 2002ContentTravel}.

These methods provide coarse insights into user information needs through their past non-textual interactions, such as clicks, purchases, and likes.  However, NL queries provide a more nuanced form of user interaction that encapsulates more explicit and personalized user information needs \cite{deldjoo2024recommendation,deldjoo2024review}. This highlights a research need for exploration of query-driven travel RSs as we do in this work. 

\subsection{Query Reformulation}
The literature on LLM-based QR methods can be broadly taxonomized into \emph{keyword}, \emph{paraphrase}, and \emph{relevant answer passage} methods. 
While such QR techniques have been studied for decades --- with well-known examples being latent semantic analysis (LSA) \cite{deerwester1990indexing, dumais1988using} and pseudo relevance feedback (PRF) \cite{1971PRF, 1990PRF, 2002PRF} --- modern LLMs introduce the ability to use internalized NL knowledge for query reformulation in highly versatile ways as we discuss below.

Keyword-based methods focus on expanding the original query to include broader coverage of subtopics or relevant terms \cite{q2dq2e, genqr, korikov2024multi}. In a recent line of work, Q2E\cite{q2dq2e} experimented with different setups for keyword-level LLM-based expansion, including zero-shot vs. few-shot prompting, chain-of-thought prompting, and the incorporation of PRF. 
ProQE \cite{rashid2024progressive} utilized LLMs in a PRF setting to assess the relevance of retrieved pseudo-documents and extract keywords from them.

Paraphrase-based methods \cite{gqrgqe} and answer passage-based methods \cite{q2dq2e, query2doc, hyde} focus on enhancing the depth of the original query by either rephrasing the query to better reflect user intent (e.g., GenQR~\cite{2023GenQR}) or enriching it with information-rich relevant answers (e.g., GQR~\cite{gqrgqe}).
Additionally, several studies propose generating relevant answer passages directly using LLMs and demonstrate effectiveness in both sparse and dense retrieval, e.g., 
Query2Doc \cite{query2doc} in few-shot settings by concatenating the original query with the answer passage and Q2D \cite{q2dq2e} using various prompting setups and a similar concatenation approach. 

Current QR methods appear to focus exclusively on either expanding the breadth or depth of queries, but not simultaneously, which we previously conjectured in Section~\ref{sec:intro} is a limitation of QR methods for broad and indirect queries.  Further, 
while non-LLM-based QR methods have been explored in RSs \cite{QueryReformulationinE-CommerceSearch, bhandari2023recqr}, there is a research need to investigate the use of LLM-based methods in query-driven RSs, which is important as the information needs and intent behind NL queries in RSs can differ from standard retrieval~\cite{lops2011content, IntentRecsys2017}.


\section{Methodology}


\subsection{Query-driven Travel Recommendation}
Let \( q \) be an NL travel destination query, and let \(\mathcal{D}\) be the set of all travel destinations. Each destination \( d_i \in \mathcal{D} \) is associated with a document \(\mathcal{C}_{d_i} = \{ c_1, c_2, \ldots, c_m \}\) describing the destination $d_i$, 
where each \( c_j \) represents a paragraph within the document.  

The objective of the travel RS is to generate a ranked list \(\mathcal{S}\) that orders each \( d_i \in \mathcal{D} \) based on a scoring function. A straightforward and effective destination scoring algorithm is defined as follows:
\begin{algorithm}
\caption{Destination Scoring Algorithm}
\label{alg:dest_score}
\begin{algorithmic}[1]
\STATE $q' \gets \text{Reformulate}(q)$ 
\COMMENT{See Section \ref{qr}}
\FOR{each destination $d_i \in \mathcal{D}$}
    \STATE $\mathbf{q'} \gets \text{Encode}(q')$
    \FOR{each paragraph $c_j \in \mathcal{C}_{d_i}$}
        \STATE $\mathbf{c}_j \gets \text{Encode}(c_j)$
        \STATE $\text{score}(q',\!c_j) \gets \text{cos}(q',\!c_j)$ \COMMENT{dense} or \text{BM25}$(q',\!c_j)$ \COMMENT{sparse}
    \ENDFOR
    \STATE $\mathcal{C}_{q'} \gets$ top-$n$ paragraphs $\{ c_1, c_2, \ldots, c_n \}$ using $\text{score}(q',c_j)$
    \STATE $\text{score}(d_i) \gets \frac{1}{n} \sum_{c_j \in \mathcal{C}_{q'}} \text{score}(q',c_j)$ \COMMENT{Avg of top-$n$ scores}
\ENDFOR
\STATE $\mathcal{S}$ $\gets$ Sort destinations $d_i$ in descending order of $\text{score}(d_i)$
\end{algorithmic}
\end{algorithm}

\subsection{Query Reformulation} \label{qr}
\begin{figure}[t!]
    \centering
    \includegraphics[width=1.0\linewidth]{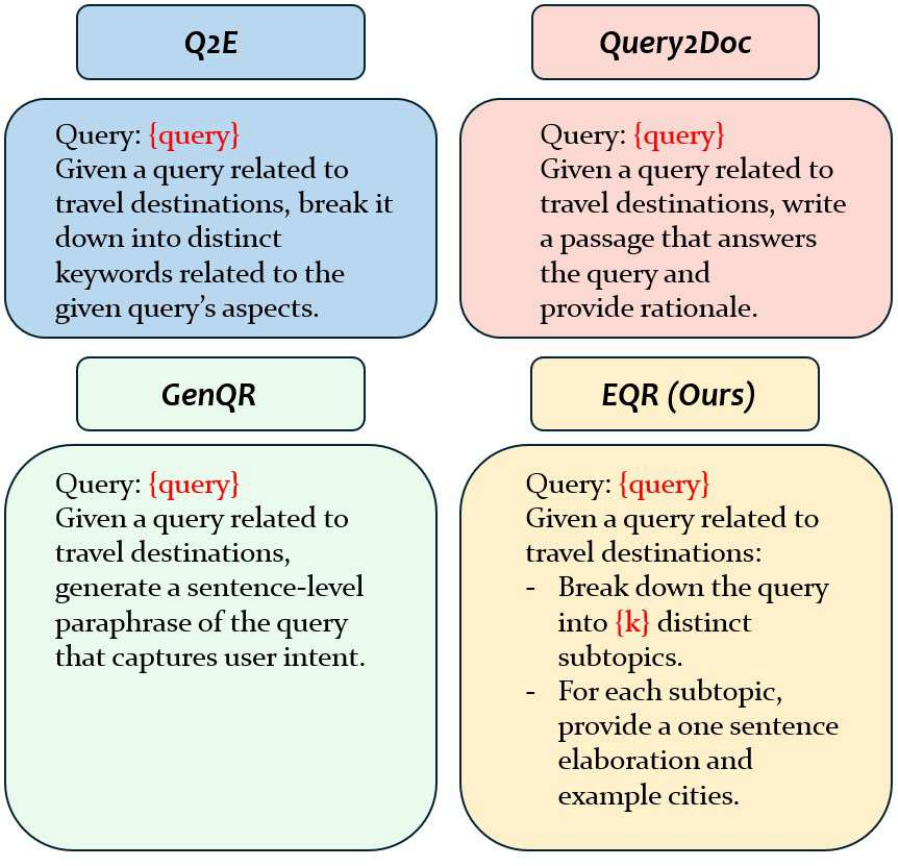}
    \caption{LLM prompts for various QR methods discussed in Section \ref{qr}, with LLM output shown in Figure \ref{fig:examples} using two broad and indirect query examples. Q2E \cite{q2dq2e} (top-left), Query2Doc \cite{query2doc} (top-right), GenQR \cite{gqrgqe} (bottom-left), and EQR (bottom-right).}
    \label{fig:prompt}
\end{figure}

In this work, we fix the structure of the Query-driven Travel Recommender as in Algorithm~\ref{alg:dest_score} while experimenting with the impact of different QR methods to implement Line 1, defined as follows:

\begin{description}
\item[No QR]\!\!: $q' = q$, which means no QR is applied.

\item[Q2E]\!\cite{q2dq2e}: $q' = q + \text{LLM}(q, \text{Q2E-prompt})$, which expands the original query by adding multiple keywords using the LLM. See Figure \ref{fig:prompt} (top-left) for a detailed prompt.

\item[Query2Doc]\!\cite{query2doc}: $q' = q + \text{LLM}(q, \text{Query2Doc-prompt})$, which generates relevant answer passages from the query using the LLM and concatenates them with the original query. See Figure \ref{fig:prompt} (top-right) for a detailed prompt.

\item[GenQR]\!\cite{2023GenQR}: $q' = q + \text{LLM}(q, \text{GenQR-prompt})$, which paraphrases the original query using the LLM. See Figure \ref{fig:prompt} (bottom-left) for a detailed prompt.

\item[EQR]\!\!: $q' = q + \text{LLM}(q, \text{EQR-prompt})$, which generates $k$ subtopic elaboration paragraphs from the query using the LLM. See Figure \ref{fig:prompt} (bottom-right) for a detailed prompt and Section \ref{eqr} for a detailed discussion on \textbf{EQR}.
\end{description}

\subsection{EQR: Elaborative Subtopic Query Reformulation} \label{eqr}
The general idea behind our novel contribution of \textbf{EQR} as motivated in Section~\ref{sec:intro} is to infer multiple subtopics from an original query (i.e., breadth) while elaborating each with information-rich content using the LLM's general reasoning abilities (i.e., depth). 

Specifically, \textbf{EQR} begins with generating $k$ distinct subtopics from a given NL query $q$, which adds a breadth aspect to capture a wider range of relevant or latent subtopics compared to answer-based and paraphrase-based methods \cite{hyde, query2doc, q2dq2e, gqrgqe}. Following this, \textbf{EQR} produces an information-rich elaboration for each subtopic, denoted $e_1, e_2, \cdots, e_k$, which adds depth via more detailed descriptions and logically entailed connections between the queries and inferred subtopics compared to keyword-based methods \cite{q2dq2e, genqr}.

\textbf{EQR} uses the LLM prompt in Figure \ref{fig:prompt} (bottom-right). The reformulated query $q'$ is a keyword merge $q' \! = \! \text{concat}(q, e1, e2, \cdots, e_k)$ for {\bf sparse retrieval via BM25} ~\cite{robertson1994bm25}, or a [SEP]-delimited text concatenation $q' \! = \! \text{concat}(q, \text{[SEP]}, e_1, \cdots, \text{[SEP]}, e_k)$ for LLM encoding and {\bf dense retrieval via cosine similarity} (cf. Line 6 in Alg~\ref{alg:dest_score}).

\section{TravelDest: Benchmark Dataset for Travel Destination Recommender Systems}

To facilitate our empirical comparison of QR methods for Query-driven Travel RS, we introduce \textbf{TravelDest}\footnote{dataset: \url{https://github.com/YifanLiu2/ROEGEN-RecSys-24-EQR.git}}
designed specifically for travel RS with a focus on broad and indirect travel queries. \textbf{TravelDest} consists of 50 broad and indirect NL travel queries, spanning various categories, including cultural, adventure, nature, entertainment, and culinary. The dataset contains full per-query labels for 774 target destination cities accessible by major airlines, each with a \texttt{WikiVoyage}\footnotemark[1] CC-licensed\footnote{WikiVoyage License: \url{https://creativecommons.org/licenses/by-sa/4.0/}} detailed text description. 

The ground truth for each query was established by asking three human labellers to manually assess all 774 target cities per query and assign relevance scores on a scale from 1 to 5. Cities achieving an average score of at least 3 were selected and verified by two additional independent travel experts to ensure relevance.


\begin{table*}[htbp]
    \centering
    \caption{Comparative performance of QR methods using different retrievers on the TravelDest dataset, averaged over 50 broad and indirect queries. An asterisk (*) denotes a statistically significant improvement (paired \(t\)-test, \(p < 0.01\)) compared to the best-performing baseline.  Error bars provide 95\% confidence intervals on the reported mean metrics.}
    \tiny

    \resizebox{\linewidth}{!}{
    \begin{tabular}{llccccc}
        \toprule
        & \textbf{QR Methods} & MAP@30 & MAP@50 & Recall@30 & Recall@50 & R-Precision \\
        \midrule
        \multirow{5}{*}{\textbf{Dense - \texttt{TAS-B}}} & \textbf{No QR} & 0.513$\pm$0.068 & 0.477$\pm$0.064 & 0.175$\pm$0.031 & 0.267$\pm$0.043 &            0.343$\pm$0.044 \\ 
         & \textbf{GenQR} & 0.510$\pm$0.065 & 0.471$\pm$0.059 &  0.175$\pm$0.031 & 0.27$\pm$0.042 & 0.351$\pm$0.042 \\ 
         & \textbf{Q2E} & 0.601$\pm$0.057 & 0.558$\pm$0.053 &  0.209$\pm$0.030 & 0.306$\pm$0.036 &  0.396$\pm$0.035 \\ 
         & \textbf{Query2Doc} & 0.699$\pm$0.058 & 0.641$\pm$0.056 & 0.209$\pm$0.034 & 0.335$\pm$0.043 & 0.414$\pm$0.040 \\ 
         & \textbf{EQR} (Ours) & \textbf{0.704}$\pm$0.055 & \textbf{0.661}$\pm$0.052 * &  \textbf{0.260}$\pm$0.036 *& \textbf{0.366}$\pm$0.044 *&  \textbf{0.455}$\pm$0.040 *\\ 

        \midrule
        \multirow{5}{*}{\textbf{Dense - \texttt{MiniLM}}} & \textbf{No QR} & 0.528$\pm$0.067 & 0.496$\pm$0.062 & 0.193$\pm$0.033 & 0.287$\pm$0.043 &            0.369$\pm$0.044 \\
         & \textbf{GenQR} & 0.554$\pm$0.061 & 0.517$\pm$0.059 &  0.197$\pm$0.033 & 0.286$\pm$0.044 &  0.369$\pm$0.044 \\ 
         & \textbf{Q2E} & 0.582$\pm$0.055 & 0.545$\pm$0.051 &  0.206$\pm$0.029 & 0.293$\pm$0.039 & 0.381$\pm$0.039 \\  
         & \textbf{Query2Doc} & 0.648$\pm$0.053 & 0.595$\pm$0.049 &  0.211$\pm$0.033 & 0.299$\pm$0.038 & 0.376$\pm$0.034 \\ 
         & \textbf{EQR} (Ours) & \textbf{0.674}$\pm$0.049 & \textbf{0.636}$\pm$0.047 * &  \textbf{0.248}$\pm$0.034 * & \textbf{0.351}$\pm$0.043 * &  \textbf{0.437}$\pm$0.036 * \\ 

        \midrule
        \multirow{5}{*}{\textbf{Sparse - \texttt{BM25}}} & \textbf{No QR} & 0.435$\pm$0.069 & 0.405$\pm$0.064 &       0.146$\pm$0.031 & 0.208$\pm$0.038 & 0.278$\pm$0.040 \\
         & \textbf{GenQR} & 0.431$\pm$0.054 & 0.390$\pm$0.050 & 0.134$\pm$0.025 & 0.204$\pm$0.034 & 0.268$\pm$0.037 \\ 
         & \textbf{Q2E} & 0.434$\pm$0.056 & 0.402$\pm$0.051 &  0.146$\pm$0.023 & 0.215$\pm$0.030 & 0.281$\pm$0.038 \\ 
         & \textbf{Query2Doc} & 0.413$\pm$0.068 & 0.387$\pm$0.061 & 0.135$\pm$0.029 & 0.209$\pm$0.039 & 0.276$\pm$0.041 \\ 
         & \textbf{EQR} (Ours) & \textbf{0.447}$\pm$0.061 & \textbf{0.411}$\pm$0.056 &  \textbf{0.147}$\pm$0.031 & \textbf{0.225}$\pm$0.041 & \textbf{0.297}$\pm$0.044 \\
        \bottomrule
    \end{tabular}}
    \label{tab:performance_metrics}
\end{table*}

\section{Experiments} 
We comparatively evaluate the query-driven RS methodology of Algorithm~\ref{alg:dest_score} using each of the QR methods defined in Section~\ref{qr} on the \textbf{TravelDest} benchmark dataset.  We aim to address the following research questions:
\begin{description}
    \item[RQ1]\!\!: How does \textbf{EQR} perform compared to other baseline QR methods?
    \item[RQ2]\!\!: How do dense and sparse retrieval methods affect the performance of each QR method in query-driven travel RSs?
\end{description}

\subsection{Setup} \label{exp}
We tested \textbf{dense retrieval via cosine similarity} using the \texttt{TAS-B} \cite{hofstatter2021efficiently} and \texttt{MiniLM}~\cite{wang2020minilm} encoders, which are both popular BERT-based sentence transformer models \cite{reimers2019sentence} based on a HuggingFace implementation~\cite{huggingface2024sentencetransformers}.  We tested \textbf{sparse retrieval via BM25}~\cite{robertson1994bm25} using the Python Rank-BM25 library implementation~\cite{rank_bm25}. For each query, the system retrieves the top-$n$ most relevant paragraphs from each \texttt{WikiVoyage} destination document.  We refer to the Appendix for full details and analysis of hyperparameter tuning for $n$.

We compared \textbf{EQR} against various LLM-based QR methods (cf. Section \ref{qr}), with both sparse and dense retrievers as outlined above. Across all QR methods, we utilized \texttt{GPT-4o} \cite{achiam2023gpt} as the commonly shared gold standard LLM for query reformulation. 

\subsection{Metrics}




We primarily focus on Recall metrics to ensure the minimal number of relevant items are missed (critical for geographical fairness) and report Recall@30 and Recall@50. Additionally, we report R-Precision to verify system precision relative to the actual ground truth size and MAP@30 and MAP@50 to assess system ability to rank more relevant destinations earlier in the results list.

\subsection{Results} \label{res}
Table \ref{tab:performance_metrics} presents the full comparative results for different QR methods on 50 broad and indirect queries using the \textbf{TravelDest} dataset, with various recall and precision evaluation metrics reported. 

\vspace{1mm}
\noindent \textbf{RQ1}: \textbf{EQR} outperforms other LLM-based QR methods across all evaluation metrics and retriever types (and with high statistical significance in many cases). \textbf{EQR} enhances coverage of retrieved destinations compared to depth-focused answer-passage-based methods, as reflected by its notable improvement at higher k-levels (i.e., @50) over \textbf{Query2Doc}, and also appears to provide better intent reasoning compared to breadth-focused expansion-based methods, as indicated by its significant improvement compared to \textbf{Q2E}.

\vspace{1mm}
\noindent \textbf{RQ2}: All QR methods achieved greater scores in dense retrieval than in sparse retrieval, as well as greater improvements in scores when compared to \textbf{No QR}. This suggests that for query-driven RSs in the travel domain (with broad and indirect queries), dense retrieval may be more effective than sparse retrieval, likely because keyword matching in sparse retrieval cannot fully capture the nuances embedded in both original and reformulated queries. Interestingly, \textbf{EQR} still performs the best among all baselines in sparse retrieval, indicating that the idea of combining breadth and depth in query reformulation applies to both sparse and dense retrieval.

\section{Conclusion}
We introduced Elaborative Subtopic Query Reformulation (\textbf{EQR}), an LLM-based QR method that adds both depth and breadth by generating multiple, information-rich subtopic elaborations to a broad or indirect query. We also introduced the \textbf{TravelDest} benchmark dataset to evaluate
query-reformation in travel RSs, with \textbf{EQR} demonstrating state-of-the-art performance across evaluation metrics and retriever types. Future work includes extending \textbf{EQR} to \textit{conversational} RSs (e.g., \cite{austin2024bayesian, friedman2023leveraging, wang2022unicrs,  li2018conversational}) and diverse travel data types such as destination reviews (e.g., \citep{kemper2024retrieval,abdollah2023self,li2023personalized}).


\bibliography{sample-ceur}

\newpage
\appendix
\section{Analysis of Hyperparameters}

\subsection{Effect of Top-n Paragraphs}

\begin{figure}[htbp]
    \centering
    \begin{subfigure}
        \centering
        \includegraphics[width=1.0\linewidth]{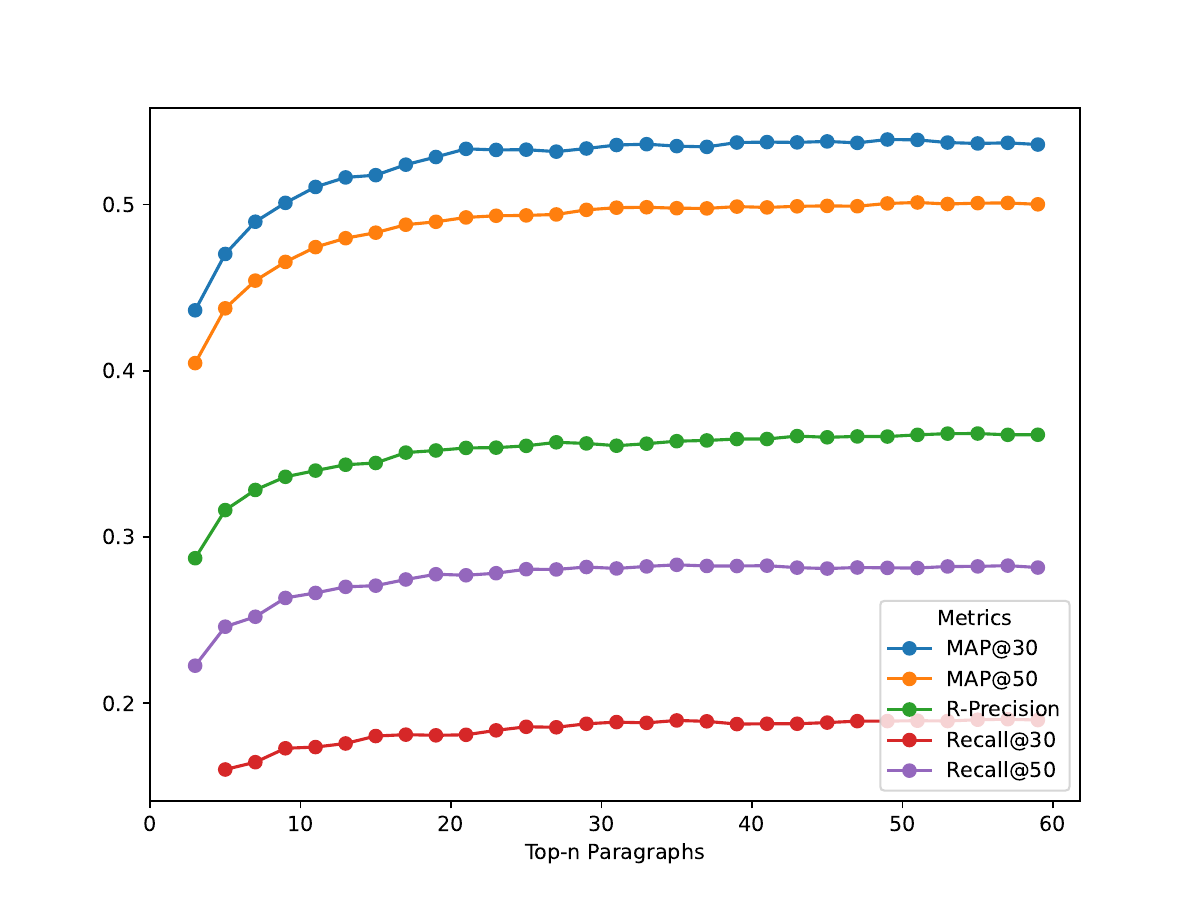}
    \end{subfigure}
    \centering
    \begin{subfigure}
        \centering
        \includegraphics[width=1.0\linewidth]{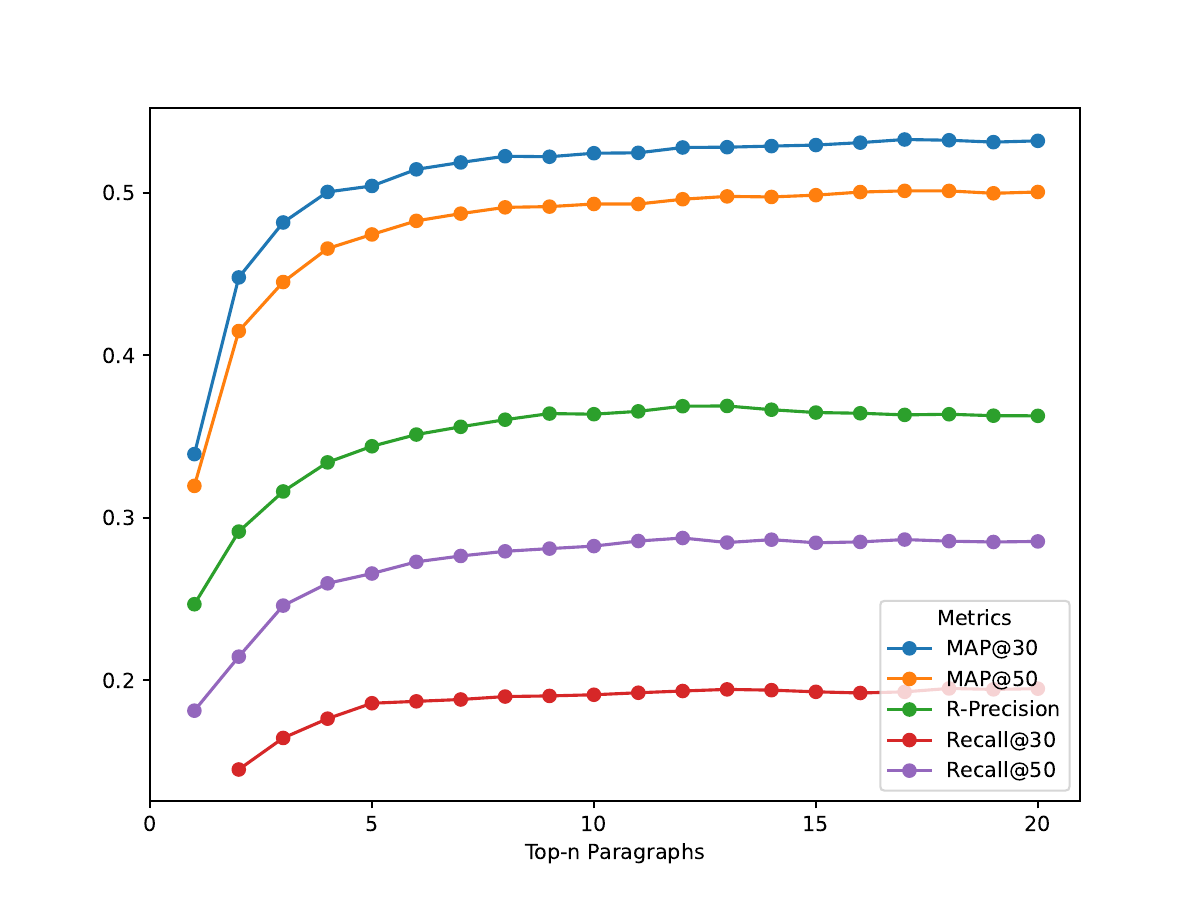}
    \end{subfigure}
    \centering
    \begin{subfigure}
        \centering
        \includegraphics[width=1.0\linewidth]{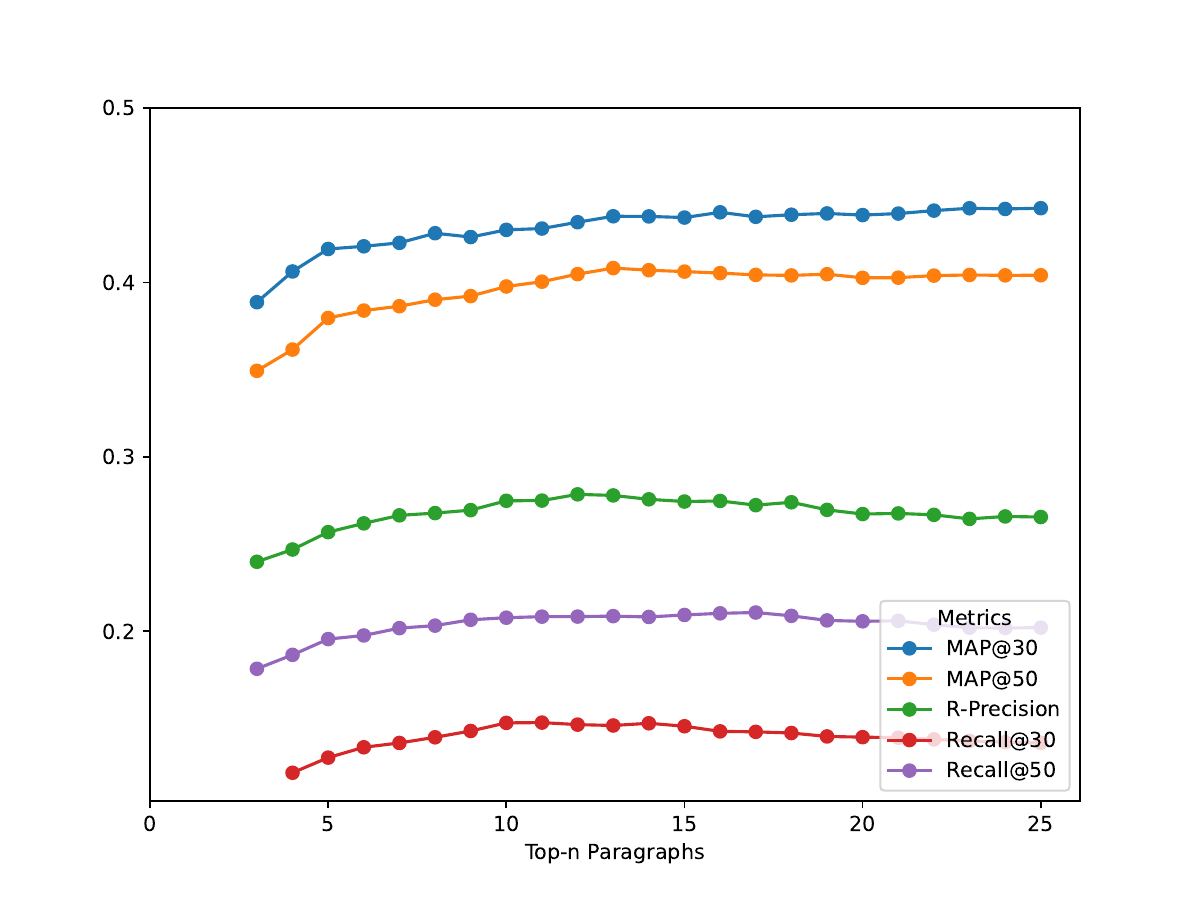}
    \end{subfigure}
    \caption{Effect of top-n paragraphs across different retrievers: Dense - \texttt{TAS-B} (top), Dense - \texttt{MiniLM} (middle), Sparse - BM25 (bottom)}
    \label{hypern}
\end{figure}

In this section, we discuss how the number of top paragraphs aggregated by the retrievers affects the performance of travel RSs in order to determine the optimal $n$ for various retrievers. We examine the scenario where no QR method is applied. According to Figure \ref{hypern}, we observed a performance increase across different evaluation metrics for all types of retrievers w.r.t. the number of top paragraphs, which was then followed by a plateau. We report the optimal values for $n$ and adopt them in Section \ref{exp}:

\begin{itemize}
    \item {\bf Dense -- TAS-B}: $n=31$
    \item {\bf Dense -- MiniLM}: $n=18$
    \item {\bf Sparse -- BM25}: $n=13$
\end{itemize}

\subsection{Effect of Number of Subtopics}

\begin{figure}[htbp]
    \centering
    \begin{subfigure}
        \centering
        \includegraphics[width=1.0\linewidth]{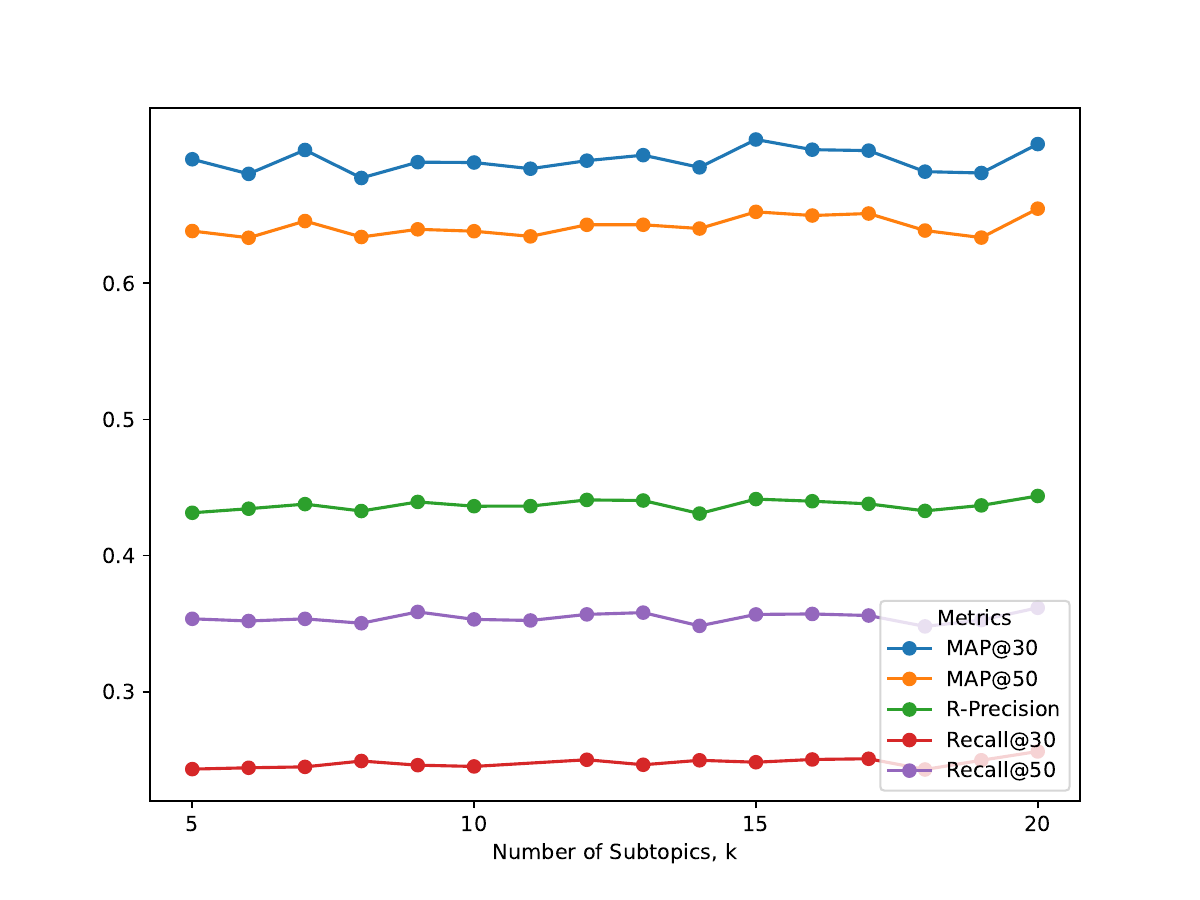}
    \end{subfigure}
    \centering
    \begin{subfigure}
        \centering
        \includegraphics[width=1.0\linewidth]{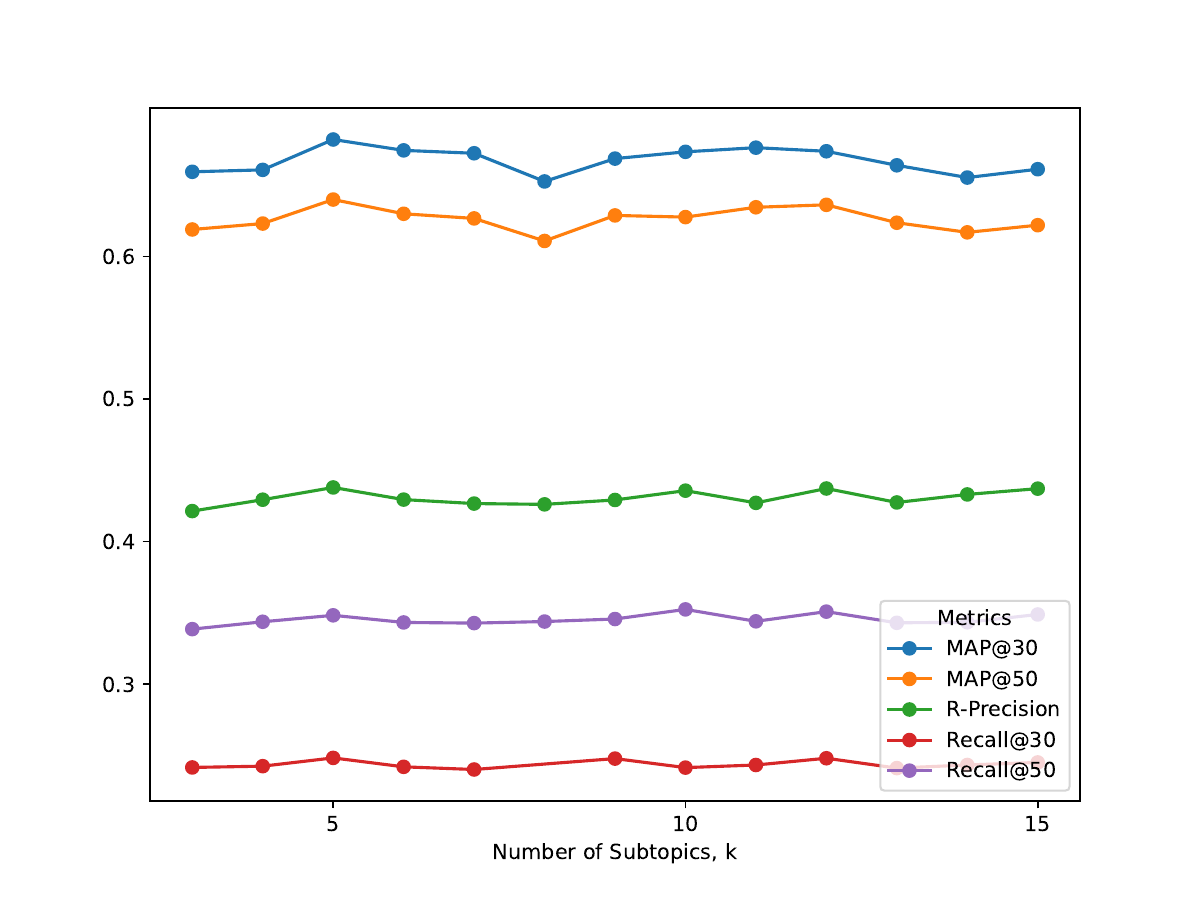}
    \end{subfigure}
    \caption{Effect of the number of subtopics in EQR across different retrievers: Dense - \texttt{TAS-B} (top) and Dense - \texttt{MiniLM} (bottom)}
    \label{hypereqr}
\end{figure}

In this section, we discuss the influence of the number of subtopics $k$ generated in \textbf{EQR} methods by testing $k$ values from 5 to 20. As shown in Figure \ref{hypereqr}, we did not observe a clear trend between $k$ and the performance of \textbf{EQR} across different types of retrievers. We report the performance for $k=12$ for all types of retrievers.

\section{Ablation Studies}
To test the robustness of \textbf{EQR}, we conducted two ablation studies on the prompts of \textbf{EQR} using the \texttt{TAS-B} dense encoder.

\subsection{Few-shot Examples}
We evaluated the performance of \textbf{EQR} with and without few-shot examples. The results did not show a significant difference, with the prompt using few-shot examples performing slightly better on recall-based metrics, while the version without few-shot examples performed better on MAP.

\begin{table}[htbp]
\centering
\begin{tabular}{lcc}
\toprule
 & \textbf{w/o few-shots} & \textbf{w/ few-shots} \\ 
\midrule
\textbf{MAP@30}           & 0.708                  & 0.704                 \\ 
\textbf{MAP@50}           & 0.662                  & 0.661                 \\ 
\textbf{Recall@30}       & 0.255                  & 0.260                 \\ 
\textbf{Recall@50}        & 0.359                  & 0.366                 \\ 
\textbf{R-Precision}        & 0.446                  & 0.455                 \\ 
\bottomrule
\end{tabular}
\caption{Comparative performance of \textbf{EQR} w/ and w/o few-shot examples using the \texttt{TAS-B} retriever}
\label{tab:ablation_study_1}
\end{table}

\subsection{Target Destinations List}
We evaluated the performance of \textbf{EQR} with and without the 774 target destination list provided in the prompt to define a valid range of choices for the LLM to select as examples. The results were mixed: the version with the target destination list performed better on MAP and Recall@30, while the version without it performed better on Recall@50.

\begin{table}[htbp]
\centering
\begin{tabular}{lcc}
\toprule
              & \textbf{w/o city list} & \textbf{w/o city list} \\
\midrule
\textbf{MAP@30}           & 0.701                            & 0.704                           \\ 
\textbf{MAP@50}           & 0.653                            & 0.661                           \\ 
\textbf{Recall@30}        & 0.256                            & 0.260                           \\ 
\textbf{Recall@50}        & 0.373                            & 0.366                           \\ 
\textbf{R-Precision}          & 0.455                            & 0.455                           \\
\bottomrule
\end{tabular}
\caption{Comparative performance of \textbf{EQR} w/ and w/o target destination list using the \texttt{TAS-B} retriever}
\label{tab:ablation_study_2}
\end{table}

\section{Examples}
In this section, we provide additional examples of various QR methods and results as shown in Tables~\ref{tab:query_example1},\ref{tab:query_example2},\ref{tab:query_example3},\ref{tab:query_example4}, and \ref{tab:query_example5}.

\clearpage

\begin{table*}[h!]
\centering
\resizebox{\linewidth}{!}{
\small 
\begin{tabular}{|p{2cm}|p{12cm}|}
\hline
\textbf{Method} & \textbf{Reformulation / Results} \\
\hline
Q2E & extreme sports; hiking trails; rock climbing; water sports; skydiving\\
\textbf{Top Results} & \textit{Galapagos Islands, Queenstown (New Zealand), Cape Town, Denver, Hong Kong, \dots}\\
\hline
GQR & Destinations known for offering a wide range of adventurous experiences \\
\textbf{Top Results} & \textit{Rio de Janeiro, São Paulo, Washington, D.C., Albuquerque, Charlotte, \dots} \\
\hline
Q2D & \underline{1. Queenstown, New Zealand}: Known as the 'Adventure Capital of the World,' offering everything from bungee jumping and skydiving to skiing and snowboarding \underline{2. Interlaken, Switzerland}: This city is a hub for adventure sports, offering activities like paragliding, skydiving, and ice climbing. \underline{3. Banff, Canada}: Located in the heart of the Canadian Rockies, Banff is a paradise for outdoor enthusiasts.\\
\textbf{Top Results} & \textit{Queenstown (New Zealand), New York City, Auckland, Santiago, Dunedin, \dots} \\
\hline
\textbf{EQR} (k=3) & \underline{Mountain Adventures} - Cities that offer hiking, climbing, and skiing opportunities in nearby mountain ranges such as Queenstown (New Zealand) and Aspen. \underline{Water Sports} - Coastal cities known for exceptional surfing, diving, and sailing activities such as Honolulu and Gold Coast. Jungle Expeditions - Locations that provide guided tours and expeditions into dense jungles and rainforests such as Manaus and Belize City. \underline{Desert Safaris} - Cities that offer desert experiences, including dune bashing and camel rides such as Dubai and Abu Dhabi. Extreme Sports - Cities that host a range of extreme sports from bungee jumping to paragliding such as Queenstown (New Zealand) and Interlaken. \\
\textbf{Top Results} & \textit{Queenstown (New Zealand), Puerto Vallarta, Tenerife, Santiago,
Gold Coast, \dots} \\
\hline
\end{tabular}}
\caption{Query: Top cities for adventure seekers}
\label{tab:query_example1}
\end{table*}

\begin{table*}[h!]
\centering
\resizebox{\linewidth}{!}{
\small 
\begin{tabular}{|p{2cm}|p{12cm}|}
\hline
\textbf{Method} & \textbf{Reformulation / Results} \\
\hline
Q2E & youth-friendly activities; budget accommodations; group tours; adventure parks; cultural experiences\\

\textbf{Top Results} & \textit{Aarhus,
        San Francisco,
        San Diego,
        Washington, D.C.,
        Rochester (New York), \dots}\\
\hline
GQR & Cities suitable for celebrating a high school graduation with friends. \\
\textbf{Top Results} & \textit{Shanghai, San Francisco, New York City, Milan, Philadelphia, \dots} \\
\hline
Q2D & \underline{1. New York City, USA}: Iconic landmarks like Times Square and Central Park offer vibrant learning and celebration opportunities. \underline{2. London, England}: Rich in history, London offers educational visits to sites like the British Museum and Tower of London. \underline{3. Tokyo, Japan}: Tokyo blends tradition and futurism, offering unique experiences in culture, technology, and cuisine.
\\
\textbf{Top Results} & \textit{New York City, London, Manchester, Philadelphia, Rome, \dots
} \\
\hline
\textbf{EQR} (k=3) & \underline{Adventure Activities}: Cities that offer exciting outdoor activities and adventures ideal for energetic young adults, such as Queenstown and Interlaken. \underline{Cultural Hotspots}: Cities rich in cultural experiences, museums, and historical sites, providing educational value, such as Rome and Athens. \underline{Beach Destinations}: Popular coastal cities with vibrant beach scenes and nightlife suitable for young travelers, such as Miami and Cancun.\\
\textbf{Top Results} & \textit{San Francisco, Milan, Shanghai, New York City, Athens, \dots
} \\
\hline
\end{tabular}}
\caption{Query: Cities for a high school graduation trip}
\label{tab:query_example2}
\end{table*}

\begin{table*}[h]
\centering
\resizebox{\linewidth}{!}{
\small 
\begin{tabular}{|p{2cm}|p{12cm}|}
\hline
\textbf{Method} & \textbf{Reformulation / Results} \\
\hline
Q2E & wellness centers; yoga retreats; meditation centers; hermal baths; sunshine beaches; nature reserve; peaceful countryside \\
\textbf{Top Results} & \textit{Palm Springs,
        Aruba,
        Cape Town,
        Gold Coast, \dots} \\
\hline
GQR & Destinations known for their tranquil environments, wellness centers, and natural beauty, ideal for a relaxing and restorative getaway.\\
\textbf{Top Results} & \textit{London,
        San Francisco,
        Shanghai,
        Rochester (New York),
        San Diego}, \dots\\
\hline
Q2D & \underline{1. Sedona, Arizona, USA}: Famous for its red rock scenery and arts community, Sedona also offers wellness retreats and spas. \underline{2. Bali, Indonesia}: Known for its beaches, rice terraces, and tranquil atmosphere, Bali is perfect for relaxation. \underline{3. Kyoto, Japan}: Renowned for temples, tea houses, and gardens, Kyoto provides a peaceful escape from daily life.\\
\textbf{Top Results} & \textit{Santo Domingo,
        São Paulo,
        Tenerife,
        Jakarta,
        Cabo San Lucas, \dots
} \\
\hline
\textbf{EQR} (k=3) & 
\underline{Spa Retreats} - Cities known for offering luxurious spa services that combine relaxation with a variety of wellness treatments, perfect for rejuvenating the mind and body, such as Budapest and Bali. \underline{Nature Escapes} - Destinations surrounded by stunning natural landscapes, ideal for outdoor activities like hiking and sightseeing, and offering a peaceful break from the fast pace of life, such as Asheville and Queenstown (New Zealand). \underline{Beachfront Relaxation} - Cities with serene and picturesque beaches, perfect for enjoying sunbathing, swimming, and rejuvenating by the sea, such as Maldives and Honolulu.\\
\textbf{Top Results} & \textit{Mombasa,
        Santo Domingo,
        Aruba,
        Puerto Vallarta,
        Maldives, \dots} \\
\hline
\end{tabular}}
\caption{Query: Cities for a rejuvenating retreat}
\label{tab:query_example3}
\end{table*}

\begin{table*}[h]
\centering
\resizebox{\linewidth}{!}{
\small 
\begin{tabular}{|p{2cm}|p{12cm}|}
\hline
\textbf{Method} & \textbf{Reformulation / Results} \\
\hline
Q2E & quaint villages; cobblestone streets; local markets; artisan shops; scenic views; historic downtown; peaceful retreats; cultural festivals; bed and breakfasts; picturesque landscapes \\
\textbf{Top Results} & \textit{Albuquerque,
        Aurangabad,
        Aarhus,
        Ottawa,
        George Town (Malaysia), \dots} \\
\hline
GQR & Quaint towns that exude charm, perfect for a peaceful and immersive getaway.\\
\textbf{Top Results} & \textit{London,
        Shanghai,
        Birmingham,
        New Orleans,
        Rochester (New York), \dots}\\
\hline
Q2D & \underline{1. Bruges, Belgium}: Famous for its medieval architecture, cobbled streets, and canals, Bruges offers a charming mix of history, cozy cafes, and art museums. \underline{2. Hallstatt, Austria}: Nestled between a lake and mountains, Hallstatt is renowned for its breathtaking landscapes and ancient salt mines. \underline{3. Carmel-by-the-Sea, California, USA}: Known for its fairy-tale cottages, art galleries, and beaches, Carmel is perfect for relaxation in a creative setting.
\\
\textbf{Top Results} & \textit{Amsterdam,
        Lisbon,
        Brussels,
        Tallinn,
        Aarhus, \dots
} \\
\hline
\textbf{EQR} (k=3) & \underline{Historic Charm}: Towns that provide a rich sense of history, featuring well-preserved architecture and deep-rooted local traditions, perfect for cultural exploration, such as Bathurst (New Brunswick) and Ljubljana. \underline{Natural Beauty}: Small towns nestled in breathtaking natural surroundings, offering opportunities for outdoor activities like hiking, photography, and nature walks, such as Aspen and Queenstown (New Zealand). \underline{Cultural Festivals}: Towns renowned for their distinctive local festivals, giving visitors an authentic insight into regional culture and traditions, such as Edinburgh and Pamplona.
\\

\textbf{Top Results} & \textit{Riga,
        Aarhus,
        Albuquerque,
        Edmonton,
        Montevideo, \dots
} \\
\hline
\end{tabular}}
\caption{Query: Charming small town cities}
\label{tab:query_example4}
\end{table*}

\begin{table*}[h]
\centering
\resizebox{\linewidth}{!}{
\small 
\begin{tabular}{|p{2cm}|p{12cm}|}
\hline
\textbf{Method} & \textbf{Reformulation / Results} \\
\hline
Q2E & off-the-beaten-path; secluded; quiet towns; remote; less touristy; undiscovered; peaceful; small towns; hidden gems; tranquil \\
\textbf{Top Results} & \textit{São Paulo,
        Manchester,
        Brussels,
        Ibiza,
        Nice, \dots} \\
\hline
GQR & Cities known for their tranquil atmosphere and less tourist traffic. \\
\textbf{Top Results} & \textit{London,
        Shanghai,
        Paris,
        San Francisco,
        Buenos Aires,
        New York City, \dots}\\
\hline
Q2D & \underline{1. Ljubljana, Slovenia}: Known for its green spaces, pedestrian-friendly streets, and relaxed vibe, ideal for a quieter European experience. \underline{2. Luang Prabang, Laos}: A serene town blending French colonial architecture and Buddhist temples, set in a lush, mountainous landscape. \underline{3. Reykjavik, Iceland}: Though popular, it offers chances to escape crowds by exploring nearby natural wonders like the Golden Circle and Blue Lagoon.
\\
\textbf{Top Results} & \textit{Reykjavík,
        Helsinki,
        Ljubljana,
        Aarhus,
        Tallinn, \dots
} \\
\hline
\textbf{EQR} (k=3)& \underline{Remote Locations}: Cities that are off the beaten tourist path, providing a sense of solitude and offering distinctive, memorable experiences, such as Iqaluit and Ålesund. \underline{Small Town Charm}: Smaller cities known for their peaceful streets, intimate atmosphere, and lack of large tourist crowds, making them ideal for a slower-paced getaway, such as Bathurst (New Brunswick) and Lethbridge. \underline{Nature Escapes}: Cities situated near expansive nature reserves and national parks, where visitors can easily disconnect from urban life and immerse themselves in the tranquility of the outdoors, such as Whitehorse and Aspen.
\\
\textbf{Top Results} & \textit{
        Brussels,
        Reykjavík,
        Ljubljana,
        Budapest,
        Venice, \dots
} \\
\hline
\end{tabular}}
\caption{Query: Best cities to avoid crowds}
\label{tab:query_example5}
\end{table*}

\end{document}